\documentclass[prl,twocolumn,superscriptaddress,floatfix]{revtex4-2}
\usepackage{graphicx}
\usepackage{latexsym}
\usepackage{amsmath}
\usepackage{amsfonts}
\usepackage{amssymb}
\usepackage{bm}
\usepackage{txfonts}
\usepackage[markup=default]{changes}
\newcommand{\fig}[2]{\includegraphics[width=#1]{#2}}
\def\bk{{\textbf{k}}}

\def\eAFM{{{\bm e}_\text{AFM}}}
\def\edSC{{{\bm e}_{d\text{SC}}}}
\def\edSB{{{\bm e}_{d\text{SB}}}}

\begin{document}
\title{Spontaneous emergence of altermagnetism in the single-orbital extended Hubbard model}
	
\author{Jin-Wei Dong}
\affiliation{Anhui Province Key Laboratory of Condensed Matter Physics at Extreme Conditions, High Magnetic Field Laboratory, Chinese Academy of Sciences, Hefei 230031, China}

\author{Yu-Han Lin}
\affiliation{Institute of Theoretical Physics, Chinese Academy of Sciences, Beijing 100190, China}
\affiliation{School of Physical Sciences, University of Chinese Academy of Sciences, Beijing 100049, China}

\author{Ruiqing Fu}
\affiliation{Institute of Theoretical Physics, Chinese Academy of Sciences, Beijing 100190, China}
\affiliation{School of Physical Sciences, University of Chinese Academy of Sciences, Beijing 100049, China}

\author{Xianxin Wu}
\affiliation{Institute of Theoretical Physics, Chinese Academy of Sciences, Beijing 100190, China}

\author{Gang Su}
\thanks{Corresponding author: sugang@itp.ac.cn}
\affiliation{Institute of Theoretical Physics, Chinese Academy of Sciences, Beijing 100190, China}
\affiliation{School of Physical Sciences, University of Chinese Academy of Sciences, Beijing 100049, China}
\affiliation{Kavli Institute of Theoretical Sciences, University of Chinese Academy of Sciences, Beijing 100049, China}

\author{Ziqiang Wang}
\thanks{Corresponding author: wangzi@bc.edu}
\affiliation{Department of Physics, Boston College, Chestnut Hill, MA 02467, USA}

\author{Sen Zhou}
\thanks{Corresponding author: zhousen@itp.ac.cn}
\affiliation{Institute of Theoretical Physics, Chinese Academy of Sciences, Beijing 100190, China}
\affiliation{School of Physical Sciences, University of Chinese Academy of Sciences, Beijing 100049, China}

\begin{abstract}
Altermagnetism (AM), the recently discovered third class of collinear magnetic order, is characterized by non-relativistic momentum-dependent spin-split electronic structure with compensated zero net magnetization.
It can arise from the conventional antiferromagnetism by introducing local anisotropy on the two opposite-spin sublattices, either through structural changes in local crystallographic symmetry or spontaneous emergence of local staggered orbital order from electron correlations in multi-orbital systems.
Here, we demonstrate on the two-dimensional square lattice that a $d$-wave AM can emerge spontaneously in the single-orbital extended Hubbard model, without invoking crystallographic anisotropy and multi-orbital physics.
We carry out mean-field studies on the concrete single-orbital $t$-$U$-$V$ model with $U$ and $V$ the onsite and nearest-neighbor Coulomb interactions, obtaining the mean-field ground states, analyzing their properties, and determining the phase diagram in the $U$-$V$ plane.
The $d$-wave AM with novel spin-transport behavior is found to be stabilized in a wide region of the phase diagram when the system is doped away from half-filling, actualized by the coexistence of onsite antiferromagnetic order and complex $d$-wave nearest-neighbor spin bond orders.
Our findings provide an alternative route to achieve AM and substantially expand the range of candidate AM materials.
\end{abstract}
\maketitle

\textit{Introduction.} - 
Altermagnetism (AM), the recently discovered third type of collinear magnetic order beyond the traditional ferromagnetism (FM) and antiferromagnetism (AFM), has gained tremendous attention in condensed matter physics \cite{Smejkal-ScAdv20, MazinII-pnas21,  SmejkalL-prx22, SmejkalL-prx22b, Mazin-prx22, YaoYugui-AdvFM24}. 
This unconventional magnetic order not only enriches our fundamental physical concepts \cite{MazinII-pnas21, SmejkalL-prx22, SmejkalL-prx22b, YaoYugui-AdvFM24, LiuQihang-prx22, LiuQihang-prx24, FangChen-prx24, SongZhiDa-prx24, ZhaoYuJun-prb24, Mazin-prx22, Smejkal-ScAdv20, WuCongjun-prb07} but also possesses immense potential for applications in low-dissipation spintronics and quantum information architectures \cite{Rafael-prl21, Smejkal-prx22, Shao-NC21, Smejkal-arxiv24, HeRan-prl23, YaoYugui-prl24, DaiJK-PRM24, FengZexin-NatElec22, Bhowal-prx24, Antonenko-prl25, DuanXunkai-prl25, LiuQihang-prl25, Takahashi-prb25}.
The unique spin symmetry of AM engenders a compensated zero net magnetization akin to AFM, while simultaneously hosting non-relativistic spin-splitting effects reminiscent of FM.
The discovery of AM has ignited a surge in materials exploration.
Symmetry-guided predictions have identified numerous candidates \cite{MazinII-pnas21, SmejkalL-prx22, SmejkalL-prx22b, Smejkal-prl23, kriegner-nc16, Mazin-arxiv22, Bose-prb24, Parthenios-arxiv25, Smejkal-ScAdv20, Roig-prb24, JWLiu-arxiv25, YuYue-nc25, Pupim-prl25} spanning insulating, semiconducting, metallic, and even superconducting regimes, while supporting experimental evidences \cite{krempaskyJ-Nat24, ZhuYP-Nat24, LeeSY-prl24, OsumiT-prb24, FedchenkoO-SciAdv24, ReimersS-NC24, RegmiRB-NC24, LinZH-arXiv24, YangGW-arXiv24, ZengM-arXiv24, RegmiRB-arXiv24, ZhangFayuan-arxiv25, Candelora-arxiv25, Jeong-arxiv25, Parfenov-jacs25, Qian-np25} from spin-resolved spectroscopy and magnetotransport measurements have been revealed in rutile oxides, diamond lattice materials, distorted perovskite systems, and so on.

AM has been shown to arise from the conventional AFM by introducing local anisotropy on the two opposite-spin sublattices.
As a result, the two sublattices in AM are no longer connected by real-space inversion ($\mathcal{P}$) or translation ($\tau$) symmetries, which is the hall mark of AFM, but linked by rotational or mirror ($\mathcal{R}$) symmetries.
Consequently, AM preserves a composite spin symmetry $\{C^{\bm n}_2||\mathcal{R}\}$ which ensures a compensated zero net magnetization, but breaks the combined $\mathcal{PT}$ symmetry and $\{C^{\bm n}_2||\tau\}$ spin symmetry that lifts the Kramers spin degeneracy and gives rise to the momentum-dependent spin-split electronic bands in the absence of spin-orbit coupling (SOC) \cite{SmejkalL-prx22, SmejkalL-prx22b, MazinII-pnas21, YaoYugui-AdvFM24}.
The operation $C^{\bm n}_2$ infront of the double-vertical-bar denotes the 180 degree spin rotation about $\bm n$-axis, and $\bm n$ here can be any direction in the plane perpendicular to the ordered moments.
In the existing model realizations of AM, the sublattice anisotropy either arises structurally from local crystallographic anisotropy of the material \cite{Okamoto-prb23, Brekke-prb23, Fernandes-prb24, Knap-prl24, GuoYaqian-MatTP23, Thomale-arxiv24, Roig-prb24, Valent-prb24, GuoHuaiming-arxiv25, Parfenov-jacs25}, or emerges spontaneously from electron correlations in multi-orbital systems where staggered local orbital order is developed to provide the sublattice anisotropy indispensable for AM \cite{Leeb-prl24, Capone-prb25, LiYing-arxiv24, Camerano-arxiv25}.

\begin{figure*}
	\begin{center}
		\fig{7.in}{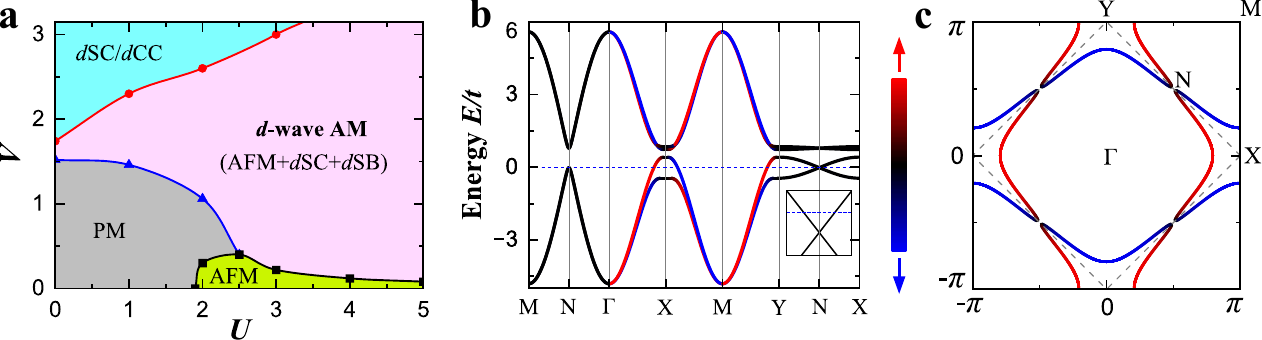} \vskip - 0.2cm
		\caption{(a) Phase diagram of the $t$-$U$-$V$ model at 8\% hole doping, consisting of PM, AFM, $d$SC/$d$CC, and $d$-wave AM metals. Solid lines denote the phase boundaries of first-order transitions. (b) Electronic structure along high-symmetry path and (c) corresponding FS of the $d$-wave AM at interactions $(U, V)$ = (3, 1.5), with the color of lines denoting the spin-splitting. Inset in (b) zooms into the band dispersion near Fermi level at $N$ point.} \label{fig1}
	\end{center}
	\vskip-0.8cm
\end{figure*}

In this work, we uncover an alternative microscopic route to AM in correlated single-orbital systems, where the required sublattice anisotropy emerges spontaneously from interaction-driven bond ordering.
Specifically, we carry out mean-field studies of the single-orbital extended Hubbard model, i.e., the $t$-$U$-$V$ model with $U$ and $V$ denoting the onsite and nearest-neighbor (nn) Coulomb interactions, on a square lattice.
The interactions $U$ and $V$ tend to drive, respectively, onsite collinear AFM and nn charge/spin bond orders at sufficient strengths.
Away from half-filling, the mean-field ground state in a wide region of the phase diagram in the $U$-$V$ plane possesses simultaneously AFM and complex $d$-wave spin bond orders.
This coexistent state engenders zero net magnetization and exhibits $d$-wave spin-splitting in the electronic structure, giving rise to the spontaneous realization of $d$-wave AM in a single-orbital correlated electron system without invoking crystallographic anisotropy and multi-orbital physics.

We note that the electronic structure of the AM realized in this work is partially spin-polarized along a direction perpendicular to the ordered moments, in sharp contrast to previously reported AM where the bands are fully spin-polarized along the ordered moments.
Furthermore, we investigate the carrier doping evolution of the coexistent state, and find that the spin-splitting on the Fermi surface (FS) is reversed between the electron-doped and hole-doped systems.
This enables a possible manipulation of spin transport via carrier doping or gate voltage, may be of interest for potential applications in spintronics. 
The $t$-$U$-$V$ model is believed to be relevant to a large class of correlated materials, our finding thus provide an alternative and plausible route to find materials with non-relativistic spin splitting, and substantially expand the range of candidate AM materials.

\textit{Model and mean-field theory.} - We start with the concrete single-orbital $t$-$U$-$V$ model on the square lattice
\begin{equation}
	H= -t\sum_{\langle ij \rangle,\alpha} \left(c^\dagger_{i\alpha} c_{j\alpha} +h.c. \right) +U\sum_i \hat{n}_{i\uparrow} \hat{n}_{i\downarrow} + V\sum_{\langle ij \rangle} \hat{n}_i \hat{n}_j, \label{tUV}
\end{equation}
where $c^\dagger_{i \alpha}$ creates a spin-$\alpha$ ($\uparrow, \downarrow$) electron at site $i$, the density operators $\hat{n}_\alpha = c^\dagger_{i\alpha} c_{i\alpha}$ and $\hat{n}_i =\sum_\alpha \hat{n}_\alpha$.
Hereinafter, we set nn hopping $t=1$ as the energy unit.
Decoupling the onsite interaction in terms of local magnetic moment $\hat{m}_i^\mu$ = $\sum_{\alpha \beta} c^\dagger_{i\alpha} \sigma^\mu_{\alpha \beta} c_{i\beta}$ $(\mu =x,y,z)$, and the inter-site interaction in terms of bonds $\hat{\chiup}^\nu_{ij}$ = $\sum_{\alpha \beta} c^\dagger_{i\alpha} \sigma^\nu_{\alpha \beta} c_{j\beta}$ $(\nu =0,x,y,z)$, with $\sigma^{\mu/\nu}$ the corresponding Pauli matrix, one reaches the mean-field Hamiltonian
\begin{align}
	H_\text{MF} = & -t\sum_{\langle ij \rangle} \left( \hat{\chiup}^0_{ij} +h.c. \right)  -{U\over 4} \sum_{i,\mu} \left[ 2m^\mu_i \hat{m}^\mu_i -\left( m^\mu_i \right)^2 \right]  \nonumber \\
	& - {V\over 2} \sum_{\langle i j\rangle, \nu} \left[ \left( \chiup^\nu_{ij} \right)^* \hat{\chiup}^\nu_{ij} +h.c. -\left| \chiup^\nu_{ij}\right|^2 \right], \label{HMF}
\end{align}
where the direct Hartree terms of the interactions depending on the electron density $n_i =\langle \hat{n}_i \rangle$ are neglected to avoid double-counting, since their contributions are already considered in density functional theory \cite{Jiang-prb16}.
The order parameters, magnetic moment $m^\mu_i =\langle \hat{m}^\mu_i \rangle$ and nn bonds $\chiup^\nu_{ij} =\langle \hat{\chiup}^\nu_{ij} \rangle$, are to be determined self-consistently by minimizing the state energy. 
We consider the quantum states to be periodic with an enlarged $\sqrt{2}\times\sqrt{2}$ unit cell, which contains 2 sites and 4 nn bonds. 
In order to obtain all possible states, we use different initial conditions for solving the self-consistent equations numerically.
When multiple states converge at a given set of interactions, we compare their state energy to determine the true mean-field ground state. 
All the low-energy states presented below are charge uniform phases.
We note that our main findings obtained in this weak-coupling mean-field theory remain valid even if the Hubbard interaction term is treated nonperturbatively by strong coupling slave-boson theory, as shown in the Supplemental Materials (SM) \cite{supp}.

\textit{Phase diagram and $d$-wave AM.} - 
The mean-field phase diagram at 8\% hole doping is presented in Fig. \ref{fig1}a in the plane spanned by $U$ and $V$.
Remarkably, in a wide regime (pink shaded area) of the phase diagram where the interactions are moderate, the electronic structure of the mean-field ground state exhibits $d$-wave spin-splitting in the momentum space.
This is clearly illustrated in the  electronic bands and corresponding FS displayed in Figs. \ref{fig1}b and \ref{fig1}c at interactions $(U, V)$ = $(3, 1.5)$, with the color of lines denoting the momentum- and band-dependent spin polarization along a particular direction to be declared below.
Furthermore, this mean-field ground state is AFM ordered with compensated zero net magnetization.
Therefore, it is a $d$-wave AM emerged spontaneously in the single-orbital $t$-$U$-$V$ model, without invoking crystallographic anisotropy and multi-orbital physics.

\begin{figure}
	\begin{center}
		\fig{3.4in}{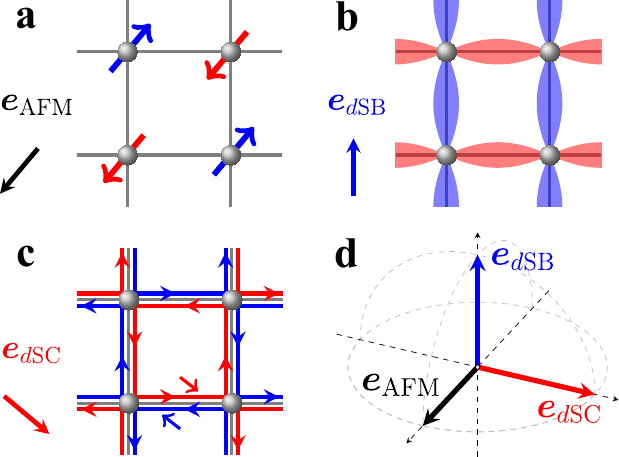}
		\caption{Schematics of (a) AFM, (b) $d$SB, and (c) $d$SC orders that coexist in the $d$-wave AM, with their spin direction along $\eAFM$, $\edSB$, and $\edSC$ respectively. The arrows on lattice sites in (a) denote the anti-aligned collinear AFM moments, the arrows along the nn bonds in (c) signal the flowing direction of spin-up and spin-down electrons, and the red and blue ellipses on the nn bonds in (b) represent the positive and negative spin hoppings. (d) The spin directions of the three orders in the $d$-wave AM are perpendicular to each other and form a right-handed chirality.} \label{fig2}
	\end{center}
	\vskip-0.8cm
\end{figure}

This $d$-wave AM is actualized by the coexistence of AFM induced by the onsite $U$ and complex $d$-wave spin bond orders driven by nn $V$.
Explicitly, the anti-aligned collinear magnetic moments ${\bm m}_i = \gamma_i (m_x, m_y, m_z) \equiv \gamma_i m \eAFM$, where $\gamma_i$ = $(-1)^{i_x+i_y}$ = $\pm 1$ on the two sublattices, as depicted in Fig. \ref{fig2}a.
The order parameter $m$ and the unit vector $\eAFM$ denote, respectively, the magnitude and the spin direction of the AFM order.
Furthermore, the complex $d$-wave spin bond orders $\chiup^\mu_{\langle ij\rangle}$ = $\eta_{ij} (\chiup'_{\mu d} +i\gamma_i \chiup''_{\mu d})$ on the nn bonds, with $\eta_{ij}$ = $(-1)^{i_y +j_y}$ the standard $d$-wave form factor.
The real components introduce the $d$-wave spin bond ($d$SB) order ${\bm \chiup}_{d\text{SB}}$ = $(\chiup'_{xd}, \chiup'_{yd}, \chiup'_{zd})$ $\equiv \chiup_{d\text{SB}} \edSB$ where spin hoppings along $x$- and $y$-axis are opposite for electrons with spins along $\edSB$ direction, as illustrated in Fig. \ref{fig2}b.
On the other hand, the imaginary components give rise to the $d$-wave spin current ($d$SC) order ${\bm \chiup}_{d\text{SC}}$ = $(\chiup''_{xd}, \chiup''_{yd}, \chiup''_{zd})$ $\equiv \chiup_{d\text{SC}} \edSC$, which generates staggered circulating currents for spins along $\edSC$ direction, as depicted in Fig. \ref{fig2}c. 
Numerical calculations \cite{supp} find that the spin directions of AFM, $d$SB, and $d$SC orders are perpendicular to each other and form a right-handed chirality $\Omega \equiv (\eAFM \times \edSC) \cdot \edSB = 1$, as shown in Fig. \ref{fig2}d. 
The order parameters $m$ = 0.269 $\mu_B$, $\chiup_{d\text{SC}}$ = 0.083, and $\chiup_{d\text{SB}}$ = 0.062 at interactions $(U, V)$ = $(3, 1.5)$.

Due to their $d$-wave symmetry, the $d$SB and $d$SC orders do not alter the Dirac point at $N$ = ($\pi/2, \pi/2$), but modify primarily the electronic structure near $X$ = $(\pi, 0)$ point.
The four eigenstates at $X$ point are given by $E_{\tau \tau'}$ = $\tau Um/2$ + $\tau' 2V (\chiup_{d\text{SC}} -\tau\Omega \chiup_{d\text{SB}})$, with $\Omega=\pm1$ for right- and left-handed chirality and the indices $\tau, \tau' =\pm 1$.
Clearly, the band splitting $4V|\chiup_{d\text{SC}}-\tau \Omega \chiup_{d\text{SB}}|$ is highly uneven on the lower ($\tau=-1$) and upper ($\tau=1$) doublets.
The $d$-wave AM with $\Omega=1$ splits primarily the lower doublet that crosses the Fermi level at $X$ point, which lowers the state energy effectively by pushing one of the band as much as possible below the Fermi level \cite{supp}, as shown in Fig. \ref{fig1}b.
Consequently, the $d$-wave AM is stabilized spontaneously as the mean-field ground state in the hole-doped  $t$-$U$-$V$ model.

We note that there are two interesting differences between the $d$-wave AM realized here and previously reported AM relying on crystallographic anisotropy \cite{Okamoto-prb23, Brekke-prb23, Fernandes-prb24, Knap-prl24, GuoYaqian-MatTP23, Thomale-arxiv24, Roig-prb24, Valent-prb24, GuoHuaiming-arxiv25, Parfenov-jacs25} or multi-orbital physics \cite{Leeb-prl24, Capone-prb25, LiYing-arxiv24, Camerano-arxiv25}.
First, the momentum-dependent spin-splitting here appears along $\edSB$, a direction perpendicular to $\eAFM$ and $\edSC$, in sharp contrast to previously reported AM where the spin-splitting is along the direction of ordered moments $\eAFM$. 
Second, the electronic bands are partially spin-polarized here, whereas fully spin-polarized in previously reported AM.
When additional SOC is considered in the latter, the electronic structure becomes partially spin-polarized as well \cite{krempaskyJ-Nat24, ReimersS-NC24, Qian-np25}.

In addition to collinear AFM and complex spin bond orders, the $d$-wave AM possesses real uniform charge bond orders $\chiup^0_{\langle ij\rangle} =\chiup'_{0s}$, which retains all the symmetries under the point group $C_{4v}$ of the square lattice, and merely introduces correction to the nn hopping integral that renormalizes the bandwidth.
In the gray shaded area of the phase diagram (Fig. \ref{fig1}a) where both onsite and nn Coulomb interactions are weak, $\chiup'_{0s}$ is the only developed order, leading to the paramagnetic (PM) mean-field ground state.
AFM phase takes over as the mean-field ground state in the green shaded area where the dominant onsite $U$ drives additional AFM order.
In the cyan shaded area with large nn $V$, the mean-field ground state becomes $d$SC as the corresponding order develops spontaneously.
While AFM order gaps out both the Dirac point at $N$ and the quadratic band touching at $X$ in the PM band structure, the $d$SC order gaps out only the quadratic band touching and gives rise to a doped Dirac semimetal \cite{Dong-prb24}.
Furthermore, the electronic bands are doubly degenerate in AFM and $d$SC, due to the existence of Kramers spin degeneracy.
Interestingly, the $d$SC phase is energetically degenerate and shares the identical band structure with the $d$-wave charge current ($d$CC) phase where the imaginary $d$-wave component $i\gamma_i\eta_{ij} \chiup''_{0d}$ develops in the charge bonds \cite{Raghu-prl08, Dong-prb24}.
It has been shown that the degeneracy can be lifted in favor of $d$SC by quantum fluctuations \cite{Raghu-prl08} or in-plane magnetic anisotropy on the mean-field level \cite{Dong-prb24}.
The $d$CC order generates circulating charge currents on the nn bonds, and produces staggered orbital magnetic flux on the square lattice.
It is also referred to as loop current and staggered flux phase studied intensively in the context of cuprates \cite{Affleck-prb88, Varma-prb97} and kagom\'e metals \cite{Feng-SciBul21, SZ-NC21, Dong-PRB23}.

\begin{figure*}
	\begin{center}
		\fig{7.in}{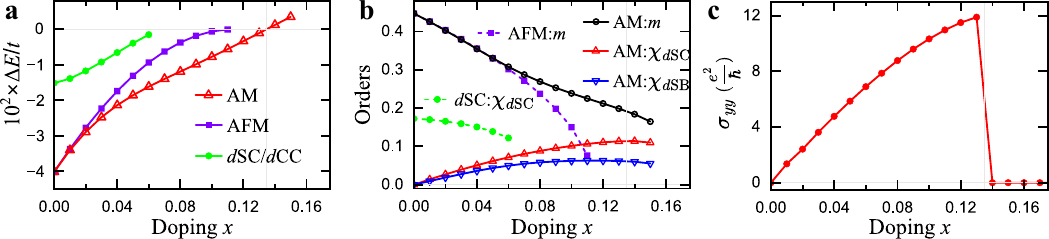}
		\caption{The hole doping $x$ dependence of (a) energy per site with respect to PM phase of the converged low-energy states, (b) the corresponding order parameters, and (c) the longitudinal spin conductivity $\sigma_{yy}$ of the mean-field ground state. The Coulomb interactions $(U, V)=(3, 1.5)$} \label{fig3}
	\end{center}
	\vskip-0.8cm
\end{figure*}

\textit{Symmetry analyses of $d$-wave AM.} - 
The symmetries of the $d$-wave AM with coexisting AFM, $d$SB, and $d$SC orders are discussed in detail in SM \cite{supp}.
It maintains the spin symmetries $\{C^\edSB_2 || M_x\}$ and $\{C^\edSC_2 || C_4\}$, thus ensures the compensate zero net magnetization of the collinear magnetic moments.
Furthermore, the coexistent state breaks the combined $\mathcal{PT}$ symmetry and spin symmetries $\{C^\eAFM_2 ||\tau\}$ and $\{C^\edSC_2 ||\tau\}$ \cite{supp}, which leads to the lifting of Kramers spin degeneracy.
According to the recent classification \cite{Cheong-npjQM25}, the coexistent state with $\mathcal{T}$-breaking is therefore a S-type strong AM which exhibits spin-split electronic bands in the absence of SOC.
We note that this AM breaks the SO(2) spin rotation symmetry and preserves the spin symmetry $\{C^\edSB_2 ||\tau\}$.
As a result, the electronic bands (Fig. \ref{fig1}b) are partially spin-polarized along $\edSB$, a direction perpendicular to $\eAFM$ and $\edSC$.
This is in sharp contrast to previously reported AM with local sublattice anisotropy where bands are fully spin-polarized along the direction of ordered magnetic moments.

The $d$-wave symmetry of the momentum-dependent spin-splitting can be illustrated by the behavior of its electronic structure under the four mirror reflections in the point group $C_{4v}$ of the square lattice, i.e., $M_x$ about $yz$-plane, $M_y$ about $zx$-plane, $M_{xy}$ about the diagonal-plane, and $M_{-xy}$ about the anti-diagonal-plane.
It preserves the spin symmetries of $\{E || M_{x,y}\tau \}$ and $\{C^\edSB_2 || M_{x,y} \}$, which implies that the electronic bands are spin-symmetric under mirror reflection $M_{x,y}$, i.e., the spin polarizations at the four momenta $(\pm \bk_x, \pm \bk_y)$ are equivalent.
Moreover, $\{C_2^\eAFM || M_{\pm xy}\}$ and $\{C_2^\edSC || M_{\pm xy} \tau\}$ symmetries in the coexistent state engender spin-antisymmetric electronic bands under mirror reflections $M_{\pm xy}$.
As a result, the electronic bands are spin-degenerate along $\Gamma$-M and X-Y lines, as shown by the spin-resolved electronic structure and FS displayed in Figs. \ref{fig1}b and \ref{fig1}c.

Interestingly, we note that the combination of any two of AFM, $d$SB, and $d$SC orders share the identical nontrivial spin space group with the coexistent state of all these three orders, provided that their spin directions are perpendicular to each other \cite{supp}.
As a result, in the self-consistent mean-field calculations we do not obtain any coexistent state with only two of these three orders, since the third order would naturally emerge to further lower the state energy.
This is also the reason why the regime of $d$-wave AM extends to $U=0$ in the phase diagram Fig. \ref{fig1}a, where kinetic AFM arises in the absence of onsite Coulomb repulsion.

\textit{Doping evolution of states and spin conductivity.} - 
Next, we fix Coulomb interactions $(U, V)=(3, 1.5)$ and monitor the evolution of states as a function of carrier doping $x$.
The $t$-$U$-$V$ model in Eq. (\ref{tUV}) with solely the nn hopping has particle-hole symmetry, we thus consider only the hole-doped side, and the states on the electron-doped side can be obtained by performing a particle-hole transformation.
The energy per site with respect to PM phase of converged low-energy states is presented in Fig. \ref{fig3}a.
If the endpoint of a curve is shown, the corresponding state becomes unstable beyond that point.
Fig. \ref{fig3}b displays the magnitudes of corresponding order parameters.
The energetically degenerate $d$SC and $d$CC states converged at $x\lesssim 0.06$ are always higher in energy, thus fail to be stabilized as the mean-field ground state.

The mean-field ground state at half filling is a fully gapped AFM insulator with the Fermi level lying in between the lower and upper degenerated doublets at $X$ point, which prevents the development of $d$SC and $d$SB orders.
However, when infinitesimal amount of holes are introduced to the system, the Fermi level is moved downward to intersect with the lower doublet at $X$ point. 
Then, within the mean-field theory, $d$SC and $d$SB can be induced by sufficient nn Coulomb interaction $V$ to lower the state energy by splitting mainly the lower doublet and pushing one of the band below Fermi level near $X$ point.
Upon hole-doping, the ordered AFM moments decrease while spin bond orders $d$SB and $d$SC increase, stabilizing the $d$-wave AM with spin-split electronic bands.
Increasing hole doping further, the mean-field ground state becomes the PM phase via a first-order transition at $x_c \simeq 0.135$, as indicated by the level crossing in Fig. \ref{fig3}a.

To demonstrate the unique spin transport properties of the $d$-wave AM, we calculate the spin conductivity $\sigma^c_{ab}$, where $c$ corresponds to the spin polarization direction of the spin current, $a$ to the flow direction of spin current, and $b$ to the direction of the applied electric field.
The $t$-$U$-$V$ model is invariant under the global SU(2) spin rotation, we thus restrict $\edSB$ and hence $c$ to be along the $\hat{z}$-axis for the simplification of numerical calculations.
Within linear response theory, spin conductivity is given by the Kubo formula \cite{Rafael-prl21,Leeb-prl24}
\begin{equation}
\sigma_{ab} = -{e\pi \over N} \lim_{\omega \rightarrow 0} \sum_{\bk,m,n} A_{n\bk}(\omega) \langle u_{n\bk} | J^z_a | u_{m\bk} \rangle A_{m\bk} (\omega)\langle u_{m\bk} | v_b | u_{n\bk} \rangle,  \nonumber
\end{equation}
where $A_{n\bk}(\omega)=-\frac{1}{\pi} [\Gamma/(\omega -\epsilon_{n\bk})^2 +\Gamma^2]$ is the band-resolved spectral function with constant scattering rate $\Gamma$ = 0.02, $v_a = \partial H_\text{MF}(\bk) /\partial \bk_a$ is the velocity operator, $J^z_a =\frac{e}{2} \{\sigma^z, v_a \}$ is the spin current operator, and $\epsilon_{n\bk}$, $u_{n\bk}$ are the eigenenergy and corresponding wave function of the Hamiltonian $H_\text{MF}(\bk)$.
The element $\sigma_{yy}=-\sigma_{xx}$ of the spin conductivity tensor is plot in Fig. \ref{fig3}c as a function of hole doping $x$.
Clearly, the longitudinal spin conductivity is zero at exactly half filling $x$ = 0, becomes nonzero and enhances gradually upon hole doping.
It reaches its maximum value at $x_c$ and then drops abruptly to zero, as the mean-field ground state undergoes a first order transition from $d$-wave AM to PM phase.

The states on the electron-doped side can be readily obtained from those on the hole-doped side by performing the particle-hole transformation, under which the magnitudes of the AFM, $d$SB, and $d$SC orders in the $d$-wave AM remain invariant whereas their spin directions are reversed.
As a result, the spin directions of these three orders form instead a left-handed chirality with $\Omega=-1$, and the band splitting at $X$ point $4V|\chiup_{d\text{SC}}-\tau \Omega \chiup_{d\text{SB}}|$ is now primarily on the upper doublet ($\tau=1$).
More interestingly, the flip of $\edSB$ is expected to reverse the direction of the spin polarization on the FS (Fig. \ref{fig1}c), and consequently lead to opposite spin conductivity on the electron- and hole-doped sides.
It thus potentially enable us to control the flowing direction of spin current by carrier doping or gate voltage, with the electric field applied along a fixed direction.
We note that, when finite next-nn hopping is introduced to the $t$-$U$-$V$ model, the particle-hole symmetry is broken, but the main findings of this work remain valid \cite{supp, Dong-prb24}, in particular the emergence of $d$-wave AM.

\textit{Summary and discussions.} - 
In this work, we have uncovered an explicit and previously unexplored microscopic realization of AM in a correlated single-orbital model.
We demonstrated explicitly on the square lattice that a $d$-wave AM emerges spontaneously in the single-orbital extended Hubbard model, due to the coexistence of onsite AFM order driven by Hubbard $U$ and complex $d$-wave spin bond orders induced by nn interaction $V$.
Unlike prior proposals of AM, this mechanism operates without crystallographic anisotropy or orbital degrees of freedom, thus substantially expands the scope of AM in correlated electronic materials.
Indeed, since the $t$-$U$-$V$ model is relevant for a diverse set of correlated materials such as cuprates and other transition metal oxides and dichalcogenides \cite{ZDWang-prb06, AHMacDonald-prl18, MMichael-prb22, LBalents-prb23, TFueno-jjap88, BKim-npjCM24, YaoH-prb24, sun-prb24, Magna-epjb98}, our findings suggest that AM could be ubiquitous in doped Mott-Hubbard systems.
For example, the stoichiometric and slightly electron-doped Sr$_2$IrO$_4$ is a canted AFM with inplane magnetic moments, and there are theoretical and experimental evidences supporting the existence of $d$SC order \cite{Torre-prl15, Zhao-np15,Zhou-prx17}.
It is thus desirable to conduct spin-resolved and angle-resolved photoemission spectroscopy measurements on Sr$_2$IrO$_4$ to examine if there is any momentum-dependent spin-splitting in its electronic structure.

Moreover, the AM discovered in this work possesses unique and novel spin-transport properties.
It exhibits partial spin-polarization along $\edSB$, a direction perpendicular to the ordered moments, which is in sharp contrast to previously reported $d$-wave AM relying on crystallographic anisotropy or multi-orbital physics where bands are fully spin-polarized along the ordered moments.
This offers an alternative route to the partially spin-polarized AM observed experimentally in the candidate materials \cite{krempaskyJ-Nat24, ReimersS-NC24, Qian-np25}, without considering SOC.
Remarkably, we found that the spin-splitting on the FS can be reversed by carrier doping, enabling tunable spin current via gating or doping with potential spintronic applications.

To this end, we note that mean-field theory is an uncontrolled variational approximation and cannot provide quantitatively reliable determination of the ground state of the extended Hubbard model.
Competing fluctuations and alternative ordering tendencies beyond the chosen mean-field ansatz may substantially modify the phase diagram.
For instance, inclusion of the Hartree channel promotes the ($\pi$, $\pi$) charge-density-wave instability at large nn interaction $V$ and significantly reduces the parameter region occupied by the $d$-wave AM phase, as demonstrated in SM. 
On the other hand, the objective of the present work is not to establish the definitive phase diagram of the $t$-$U$-$V$ model, but rather to identify and analyze a possible interaction-driven route toward AM within a concrete microscopic setting.
Therefore, despite its limitations, the mean-field framework remains useful for elucidating possible symmetry-breaking tendencies and electronic reconstruction mechanisms associated with bond-centered spin order. 
Further investigations of the extended Hubbard model using more sophisticated methods \cite{Mitsutaka1990JPSJ, Metzner2016PRL, Garnet2017Science, ZhangShiwei2022PRR, ZhangShiwei2024Science} are clearly desirable.

\textit{Acknowledgments.} -This work is supported by the National Key Research and Development Program of China (Grant No. 2022YFA1403800 and 2023YFA1407300), the National Natural Science Foundation of China (Grant No. 12374153, 12447101, and 11974362), the Innovation Program for Quantum Science and Technology (Grant No. 2024ZD0300500), and the Postdoctoral Fellowship Program of CPSF (Grant No. GZC20241749).
Z.W. is supported by the U.S. Department of Energy, Basic Energy Sciences
(Grant No. DE-FG02-99ER45747).
Z.W. thanks the Kavli Institute for Theoretical Sciences, University of Chinese Academy of Sciences for hospitality during his sabbatical visit.
Numerical calculations in this work were performed on the HPC Cluster of ITP-CAS.

\bibliography{ref}

\end{document}